\documentstyle[12pt]{article}
\begin{document}
\topmargin 0pt
\oddsidemargin 5mm
\begin{flushright}
UT-KOMABA/93-12\\
August 1993
\end{flushright}
\vskip 10mm
\headheight 0pt
\headsep 0pt
\topskip 9mm

\begin{center}

{\large

Finite $N$ Analysis of Matrix Models for $n$-Ising Spin \\
on a Random Surface }
\footnote{submitted to Physica A, dedicated to Prof. Kyoji Kawasaki
on the occasion of his retirement.}
\vspace{24pt}

{\sc  Shinobu Hikami }

\vspace{6pt}

{\sl Department of Pure and Applied Sciences, University\\
of Tokyo, Meguro-ku, Komaba 3-8-1, Tokyo 153, Japan\\}

\end{center}

\vspace{24pt}

\begin{center}

{ Abstract}
\end{center}
\vspace{6pt}
  The saddle point equation described by the eigenvalues of  $N\times N$
  Hermitian  matrices
  is analized for a finite $N$ case and
  the  scaling relation for the large $N$
  is considered.
   The critical point and
   the critical exponents of matrix model are obtained by the finite $N$
scaling.
   One matrix model and two-matrix model are
   studied in detail. Small $N$ behavior for $n$-Ising model on a random
surface
   is  investigated.
\newpage
\addtolength{\baselineskip}{0.20\baselineskip}
\begin{center}
{\large 1. Introduction}
\end{center}
\vskip 6 mm

     The large N limit of matrix models  has attracted theoretical interests
   in various fields including
    random matrix model, gauge theory and phase transition.
   Recently, it becomes clearer that the matrix model is
   closely related to the
   string field theory.  Ising model on a random surface is described by
   a two-matrix model,  in which the planar surface is
   generated in the large N limit and the up spin and down spin asigned
   by two different matrices. In this case, an external
   magnetic field on Ising spins
   can be introduced and the model is exactly solvable [1]. The critical
    exponents of the specific heat, the susceptibility and the spontaneous
   magnetization are obtained exactly, and they agree with the conformal
   field theoretical result [2].
   Their values are different from the well known Onsager solution of 2D Ising
    model on the regular lattice. The difference is caused by
    the random surface, which is interpreted as 2D quantum gravity.
    The Ising spin has c=1/2 central charge due to fermionic character,
    and consequently
    Ising model on a  random surface describes
    c=1/2 matter field coupled to 2D quantum gravity.

     The large N limit of matrix model gives planar Feyman diagrams
     in the perturbation of the coupling constant [3]. This planar diagrams
     gives us a random surface of the fixed genus.
    On each vertex of the descretized
     random surface, one can put different n-species Ising spins,
    which are independent on different species
    and only interact through random
     statistical average [4,5]. This n-Ising model on a random surface is
    particularly interesting since the matter field  has central charge
    c=n/2 which can be larger than one. Unfortunately, this model has not
    been solved yet for $n>2$ although it is described by $2^{n}$ matrix model.
    Only numerical analyses exist in this case [4,5].
     To investigate the large N limit of the matrix model, several techniques
    have been introduced. The saddle point equation [3],
     Schwinger Dyson equation,
    the orthogonal polynomial method [6], the double scaling method [7]
    and the renormalization group method [8] are considered. In this paper,
    we pursue the saddle point equation method for matrix models, since
    it  is simple and it
    gives clear interpretation for the transition. Although the large
    N limit should be taken for applying the saddle point method,
      it may be interesting to investigate
     the finite N case, and to understand
    the mechanism of the phase transition together with a
    finite N scaling behavior. We hope that finite N saddle point
   analysis becomes complemental one to the series expansion of [4].
   It is expected that our method also
   work for n-Ising case,
   and a small N calculation is presented for this as a preliminary
   study.

\newpage
\begin{center}
{\large 2. One Matrix Model}
\end{center}
\vskip 6mm
  The one matrix model has a following Hamiltonian,
\begin{equation}
H = {1\over 2}{\rm Tr} M^{2} + {g\over N}{\rm Tr} M^{4}
\end{equation}
\noindent
where $M$ is a $N\times N$ Hermitian matrix. Denoting the eigenvalues of
this matrix by $\lambda$ which are all real, the partition function is given by
\begin{equation}
Z = \int d\lambda \Pi_{i<j} (\lambda_{i} - \lambda_{j})^{2} \exp [ -{1\over 2}
\sum \lambda_{i}^{2} - {g\over N}\sum \lambda_{i}^{4}].
\end{equation}
The large N limit of this partition function is evaluated by the saddle point
equation of $\lambda$. Exponentiating the measure, and differentiating the
exponent by $\lambda_{i}$ we obtain the saddle point equation.
\begin{equation}
   - \lambda_{i} - {4g\over {N}}\lambda_{i}^{3} + \sum {2\over {\lambda_{i} -
\lambda_{j}}} = 0.
\end{equation}

 There appears
a critical value of $g$, beyond which
there is no real eigenvalue of $\lambda$.
 The free energy is expanded in the power series of $g$,
\begin{equation}
F = \sum c_{k} g^{k}
\end{equation}
\noindent
where $c_{k}$ behaves for large $k$
\begin{equation}
       c_{k} \sim A^{k}k^{ -3 + \gamma_{st} }.
\end{equation}
The series of $F$ is a convergent series for the large N limit, and $A$
is $-1/g_{c}$. For the one matrix model of (1), $g_{c}$ is -1/48 [3].
The exponent $\gamma_{st}$ is called as a string susceptibility and becomes
-1/2 for one matrix model.

Although the saddle point equation is justified for the large N limit
we apply it for a finite N matrix model. We approximate the large N
Riemann-Hilbert integral
 equation by a finite discrete eigenvalue equation.
It is easy to find the critical value of $g_{c}$ for such finite N
saddle point equation and increasing number of $N$, $g_{c}$ is
expected to become  $g_{c}$ of the large $N$ limit.
Our aim is to develop the method for obtaining the critical coupling
$g_{c}$ or cosmological constant $A$ of (5), instead of calculating the
partition function which requires the integration of the eigenvalue $\lambda$.
Our method is similar to Lipatov large order calculation, but for
finite $N$ the series of (4) becomes asymptotic and there appear k-
factorial coefficient instead of (5). Therefore, Lipatov method [9] is
not appropriate for our problem.
 We will show that this finite N saddle point method is practically useful
for obtaining the critical value $g_{c}$ and the string susceptibility
 $\gamma_{st}$.
\par
  For example in the case of N=2, we have only two eigenvalue $\lambda_{1}$ and
$\lambda_{2}$.
The effective Hamiltonian after exponentiating the measure part,
\begin{equation}
H_{eff} = {1\over 2}(\lambda_{1}^{2}+\lambda_{2}^{2}) + {g\over
2}(\lambda_{1}^{4}
+ \lambda_{2}^{4}) -  \ln (\lambda_{1}-\lambda_{2})^{2}
\end{equation}
The saddle point equation is obtained by the differentiating this Hamiltonian
by $\lambda_{1}$ and $\lambda_{2}$. There is a solution that
 two eigenvalues are symmetric around
zero,  $\lambda_{1}=-\lambda_{2}$. Thus the equations reduce to one equation,
\begin{equation}
  \lambda + 2g\lambda^{3} - {1\over \lambda} = 0
\end{equation}
This quadratic equation gives $\lambda = (-1 {\pm} (1+8g)^{1/2})/4g$ and
$g_{c}=-1/8$. The physical solution appears in + sign, which becomes
finite for $g \rightarrow 0$.
 Two solutions degenerates at $g_{c}$ and beyond this,
there is no real eigenvalue $\lambda$. The eigenvalue should be real since
a matrix $M$ is Hermitian.
\par
The obtained critical value
 $g_{c}=-1/8$ is far from -1/48, but increasing $N$, we
see that the result approaches to the correct one smoothly.
We see later that the root singularity near $g_{c}$ found in N=2 case
is a correct answer
for $N=\infty$.
 For N=3, considering $\lambda_{1}= -\lambda_{3}$,
 $\lambda_{2}=0$ solution for three eigenvalues,
we obtain $g_{c}=-1/16$. When N is even number, we take a
solution $\lambda_{1}=-\lambda_{N}$, $\lambda_{2}=-\lambda_{N-1}$,...,
and for N odd, we have $\lambda_{1}=-\lambda_{N}$,...,$\lambda_{N+1/2}=0$.
The critical value of $g_{c}$ is easily obtained numerically by
the  investigation
of the largest eigenvalue $\lambda_{1}$. The obtained critical value
 $g_{c}$ is shown
in table 1.
They are indeed approaching to -1/48.

 Since the finite N scaling is
expected, we discuss the critical exponent $\nu$ defined by
\begin{equation}
  A_{N} = A_{\infty} + {c\over N^{\nu}}
\end{equation}
where $A_{N}=-1/g_{c}$ for a finite N, and $A_{\infty}= 48$. The exponent
$\nu$ is scaling exponent defined by
\begin{equation}
  F \sim (g - g_{c})^{2-\gamma_{st}}f( N^{\nu}(g-g_{c}))
\end{equation}
Denoting the difference between $A_{\infty}$ and $A_{N}$ by $D_{N}$
\begin{equation}
  D_{N} = A_{\infty} - A_{N} \sim N^{-\nu}
\end{equation}
 we have a ratio $R_{N}=D_{N}/D_{N-1} \sim ((N-1)/N)^{\nu} \sim 1-\nu/N$
which is plotted in Fig. 1. The asymptotic coefficient of 1/N is estimated
as $\nu$ is 0.8 which is precisely same as the exact value of the
pure gravity case with $c=0$.

It is somehow remarkable that without calculating
the free energy, we obtain the correct values of $A$ and the string
susceptibility through a finite N saddle point equation up to order N=8.
 The string
susceptibility $\gamma_{st}$ is related to $\nu$ by
\begin{equation}
    \gamma_{st} + {2\over \nu} = 2,
\end{equation}
and we have $\gamma_{st}=-1/2$ since $\nu=4/5$.

 It is also interesting to study how the largest eigenvalue $\lambda_{1}$
behaves near $g_{c}$ for finite N, and to see the eigenvalue density
behaves at $g_{c}$.  It is known that the density of
the eigenvalue, which obeys usually the semi-circle law, becomes singular
  at $g_{c}$.
The density $\rho$ vanishes as [3]
\begin{equation}
  \rho \sim (\lambda - \lambda_{c})^{3\over 2}.
\end{equation}
 We find that the histogram of the density of eigenvalues for $N=8$
agrees with this behavior.
 The density
of eigenvalue at the critical value $g_{c}$ vanishes in general as
$\rho \sim (\lambda - \lambda_{c})^{-1/\gamma_{st}-1/2}$.

   The scaling between $(\lambda-\lambda_{c})$ and $(g_{c}-g)$ exists [10],
and it is given by
\begin{equation}
    \lambda - \lambda_{c} \sim ( g_{c} - g )^{-\gamma_{st}}.
\end{equation}
Since $\gamma_{st}= -1/2$ for one matrix model, this scaling relation is
satified apparently in N=2 case (7) as a solution of a quadratic equation.
 For the large value of N, also this root singularity is preserved and
we checked it up to N=8.
\vskip 20mm
\begin{center}
{\large 3. One Matrix Model with $({\rm Tr} M^{2})^{2}$ Interaction
}
\end{center}
\vskip 4mm

    It is straightfoward to apply our method to one matrix model
with $({\rm Tr} M^{2})^{2}$ interaction,
\begin{equation}
H = {1\over 2} {\rm Tr} M^{2} + {g\over N} {\rm Tr} M^{4} + {g' \over
N^{2}} ( {\rm Tr} M^{2})^{2}.
\end{equation}
where $ g'$ is an additional coupling constant. When $g=0$, this model becomes
equivalent to $O(N)$ vector model with $\gamma_{st}=1/2$ and $-1/g'_{c}
= 16$ [11].
Recently, it was found that there is a critical point at $g=-3/256$ and
$g' =-9/256$, where the string susceptibility has a positive value
$\gamma_{st} =1/3$ [12].
 It is interesting to investigate this model by our saddle point equation,
since $\gamma_{st}=1/3$ is positive and clearly (13) can not be applied.
The calculation is very easy and  we have evaluated the finite N
saddle point equation and obtained the critical values up to order N=7.
The results of the evaluation for the  critical values of $g_{c}$ and $g'_{c}
$ are given in Fig. 2. The lines of finite N solutions smoothly
converge to $n=\infty$ solution as expected. Therefore, our finite N
analysis works well for this model.

 The scaling relation of (10) is
investigated near the critical point $-1/g_{c}=3/256$ and $-1/g_{c}'=9/256$.
For fixed $-1/g_{c}'=9/256$, we evaluated $g_{c}$ from the finite N saddle
point equation. The critical value of $g_{c}$ is determined such that
 there is no real solution of the eigenvalue beyond $g_{c}$.
 In table 2, the obtained value of $g_{c}$ is represented.
Using the same ratio method $R_{N}=D_{N}/D_{N-1}$, we estimated the scaling
exponent $\nu$ in (10). We obtained for this critical point $\nu=6/5$.
Asumming (11) is still valid in this case, we obtain precisely $\gamma_{st}
= 1/3$.

   We have also verified that at a fixed $g'=-1/16$, near $g_{c}=0$,
the ratio method gives $\nu=4/3$ which leads to $\gamma_{st}=1/2$ by (11).
This point at $g=0$ and $g'=-1/16$ corresponds to the vector O(N) model
and it is known that this model has $\gamma_{st}=1/2$ [12].

  It is unexpected  result that we have a correct scaling exponent
$\nu$ by the finite N saddle point equation. Usually the scaling relation of
(9) and the exponent $\nu$ are derived by the double scaling limit [7],
based upon the orthogonal polynomial analysis for the free energy.
 We have restricted our
investigation only on the saddle point equation.

\vskip 20mm
\begin{center}
{\large 4. Two Matrix Model }
\end{center}
\vskip 4mm

   Two matrix model represents Ising model on a random surface and it is
given by
\begin{equation}
  H = {1\over 2} {\rm Tr} (M_{1}^{2} + M_{2}^{2}) - a {\rm Tr} M_{1}M_{2} +
   {g\over N} {\rm Tr} (M_{1}^{4} + M_{2}^{4})
\end{equation}
where $a$ is a coupling constant related to the nearest neighbour
spin interaction J
devided by the temperature kT, $\beta=J/kT$,,
\begin{equation}
    a = \exp( - 2\beta )
\end{equation}
 The first and the third term of the Hamiltonian are written by
the eigenvalues $\lambda$ and $\xi$ of the matrix $M_{1}$ and $M_{2}$.
For the two matrix, the Hermitian matrix is diagonalized by the unitary
matrix and it is possible to integrate this unitary matrix. Then,
partition function is written only by the eigenvalues with Haar measure  [8],
\begin{equation}
  \Xi =  C\int \Pi_{i} d\lambda_{i} a^{{-N(N-1)\over 2}}\Pi_{i<j}
 (\lambda_{i} - \lambda_{j})(\xi_{i} - \xi_{j}) \exp ( -V_{0} + a\sum
\lambda_{i}\xi_{i} ))
\end{equation}
where $V_{0}$ is the first and the third term in (15),
\begin{equation}
    V_{0} = {1\over 2}\sum (\lambda_{i}^{2} + \xi_{i}^{2} )
      +{g\over N} \sum (\lambda_{i}^{4} + \xi_{i}^{4} )
\end{equation}

 For N=2, after exponentiating the measure part, we have an effective
Hamiltonian,
\begin{equation}
  H_{eff} = (\lambda_{1}^{2} + \xi_{1}^{2}) + g (\lambda_{1}^{4} + \xi_{1}^{4})
  - \ln[{\lambda_{1}\xi_{1} \over 2} {\sinh}(2a\lambda_{1}\xi_{1})]
\end{equation}
where we use $\lambda_{1}=-\lambda_{2}$, $\xi_{1} = - \xi_{2}$. We  dropped
the irrelevant term which is vanishing in the integration of eigenvalues
before the exponentiation.
When we take the solution that
 these eigenvalues are symmetric, $\lambda_{1}=\xi_{1}$,
we have
\begin{equation}
    \lambda_{1} + 2g\lambda_{1}^{3} - {1\over {2\lambda_{1}}} -
    {a\lambda_{1} \over {{\tanh}(2a\lambda_{1}^{2})}} = 0
\end{equation}

 The critical value of $g_{c}$ is obtained from this equation.
There  is no real solution beyond $g_{c}$. We obtain for a=0, $g_{c} = -1/8$
and
$ -(1-a^{2})^{2}/g_{c} = 16$ for $a=1$. The result that the
 value of $(1-a^{2})^{2}/g_{c}$
at a=1 becomes twice of the value at a=0 is consistent with the observation
 in the perturbation of $g$ [4].
In Fig. 3, the value of $g_{c}$ is plotted as a function of $a$. Also
 the large N exact value of $g_{c}$ is given; they  consist
of two solutions [1]: the low temperature phase and the high temperature
phase respectively given by
\begin{equation}
      -(1-a^{2})^{2}/g_{c} = {48(1-a^{2})^{2} \over { 1 - {8\over 3}a^{2} } }
\end{equation}
\begin{equation}
      -(1-a^{2})^{2}/g_{c} = {18 (1 + {\sqrt{a}})^{2}(1+a)^{2} \over
   {{\sqrt{a}}(2+{\sqrt{a}})}}
\end{equation}
\break
 Taking the useful
 analogy that
 the partition
function of (17) is a grand canonical partition function with identification of
$g$
as the exponential of chemical potential $e^{\mu}$, and that the canonical
patition function is $A = -1/g_{c}$ itself, we have the free energy
of the Ising spin on a random surface as
\begin{equation}
  F(a) = - \ln A(a)
\end{equation}
  For the two-matrix model, there is a critical value of $a$, where the
spontaneous magnetization vaishes. The phase described for $a>1/4$
 corresponds to the
disorder phase (22) and the phase for $a<1/4$  is a low
 temperature ordered phase (21).

 The magnetic field $B$ for Ising spin can be introduced by the change of the
coupling $g$ in $V_{0}$ as
\begin{equation}
V_{0}(B) = {1\over 2} {\rm Tr} (M_{1}^{2} + M_{2}^{2}) + {ge^{B}\over {N}
}{\rm Tr} M_{1}^{4}
 + {ge^{-B}\over {N}} {\rm Tr} M_{2}^{4}
\end{equation}
At zero temperature a=0, the free energy $F(a,B)$ and the magnetization $M$
are
\begin{equation}
    F(a,B) = \ln g - B
\end{equation}
\begin{equation}
    M = -dF/dB = 1
\end{equation}

 For a finite magnetic field $B$, the saddle point equation gives the
asymmetric solution, $\lambda_{i} \neq \xi_{i}$. The low temperature
phase, i.e. the symmetry breaking phase also is described by  the
asymmetric solution.
 The asymmetric $\lambda \neq \xi$ solution is obtained easily,
and $g_{c}$ is plotted by the dotted line in Fig. 3.
Expanding (17) for  small $a$, we have up to order $a^{2}$,
\begin{equation}
   - {1 \over { g_{c}}} = {8 \over { 1- {8\over 3}a^{2} }}
\end{equation}
  Although it diverges at $a=1/2$ when the higher order is included,
 instead of $a^{2}=3/8$,
up to order $a^{2}$ the behavior is similar to the exact solution of (21)
which shows the divergence at $a^{2}=3/8$.

   The asymmetric solution has a larger free energy than the symmetric
one, and thus there is no symmetry breaking for N=2. The difference appears
at order $a^{4}$.  However for $a<1/4$,
the difference is very small as shown in Fig. 3, and the magnetization
evaluated for a finite B shows effectively that there is a phase transition.
We notice that it is necessary to subtract $N(N-1)/N$ terms in the expansion
of $a$ of $\exp (a\sum \lambda_{i}\xi_{i})$. Otherwise, the saddle point
solution gives a wrong answer specially for small a. It is related to the
factor
of $a^{-N(N-1)/2}$ which is divergent for $a \rightarrow 0$. Applying
the saddle point equation for the two matrix model is subtle and already
discussed in [13]. Since the first few terms in the expansion of $a$
  have no contribution in
the integral up to order $a^{N(N-1)/2}$, we safely replace $\exp(a\sum
\lambda_{i}\xi_{i})$ term by
\begin{equation}
       \sum_{k=N(N-1)/2+1}  a^{k}(\sum \lambda_{i}\xi_{i})^{k}/\Gamma(k+1)
\end{equation}
For N=5 and N=7 with the expression of (28), the symmetric solution
$\lambda_{i} = \xi_{i}$ gives
the similar curves as the exact solution of the high temperature
phase of (22). For small $a$, they indicate the divergence
 in the large N limit same as the correct expression of (22). Our (28) may be
not sufficient for deriving the correct saddle point equation, since it also
has terms which do not contribute to the integral.
Indeed for even number of $N$, the curve of $-1/g_{c}$  of the symmetric
solution becomes flat for
 small $a$.

  Since the critical value $-1/g_{c}$ should be same as one matrix model for
$a=0$,
and the low temperature phase has a continuous curve starting
 at $a=0$, if we obtain the high temperature curve which gives
divergence at $a=0$, it  concludes that there is a phase transition.

  For N=7 case, we observe that the largest eigenvalue $\lambda_{1}$ has
a maximum value at certain value of $a=0.4$,
  and below this value $\lambda_{1}$ starts to decrease.
This point $a$ seems to be a critical point. This point is  greater
than the point where the value of $-1/g_{c}$ becomes minimum. The situation
is similar to the exact solution shown in Fig.3.

 More precise small $a$ expansion using (17) is possible. Noting that
the measure $J$ is written by Vandermonde determinant in (17), we have
invariance under cyclic exchange of eigenvalues and we have antisymmetric
exchage between two different eigenvalue. The nonvanishing term in
integration of (28) at order $a^{k}$  takes a following form due to
these symmetries

\begin{equation}
    \lambda_{1}^{l_{1}}\lambda_{2}^{i_{2}}... \lambda_{N}^{l_{N}}
\end{equation}
where $ l_{1}+...+l_{N}=k$, and $l_{i} > l_{j}$ for $i<j$.
Using these nonvanishing terms in a small $a$ expansion, we obtain for N=2
as
\begin{equation}
     -{1\over g_{c}} = {8\over {1 - {8\over 3}a^{2}}} + O(a^{4})
\end{equation}
This result coincides with the previous one (27). For N=3, we find
\begin{equation}
 -{1\over g_{c}} = {16\over { 1 - 3a^{2}}} + O(a^{4})
\end{equation}

   Since the value $-(1-a^{2})^{2}/g_{c}$ at $a=1$ is twice of the value
at $a=0$ [4], we apply Pad\'e approximation of this quantity based
upon the small $a$ expansion. Before making Pad\'e analysis, we
check the validity of this approximation using the exact expression of
(21). Up to order $a^{2}$, we have
\begin{eqnarray}
   -{(1-a^{2})^{2}\over g_{c}} &=& 48 ( 1 + {2\over 3}a^{2})\nonumber\\
                               &=& 48 {{ 1 + b_{1}a^{2}}\over {1 + c_{1}a^{2}}
}.
\end{eqnarray}
where $b_{1}$ and $c_{1}$ are determined with the condition that at a=1
we have $(1+b_{1}a^{2})/(1+c_{1}a^{2})=2$. We get $b_{1}=1/3$ and $c_{1}=
-1/3$. This [1,1] Pad\'e is crude and it gives 50.04 for $a=1/4$ while
the exact value at $a=1/4$ is 50.63.

  Next order, up to order $a^{4}$, [2,1] Pad\'e approximation becomes
\begin{equation}
     -{(1-a^{2})^{2}\over {g_{c}}} = 48 {(1 + 8a^{2}+{23\over 3}a^{4})\over
{(1 + {22\over 3}a^{2})}}
\end{equation}

 This [2,1] Pad\'e gives 50.36 for $a=1/4$, and the result is improved.
The difference from the exact value is 0.5 percent. Thus we see that Pad\'e
approximation is effective. Moreover, the second derivative of the logarithm
of (33) by $a$ shows the maximum at $a=0.24$, which agrees with the exact
critical value at $a=1/4$. This means that
 we have a method of estimation of $a_{c}$ as a specific heat peak.

  Using the results of (30) and (31), we have Pad\'e approximation for
$N=2$ and $N=3$ as
\begin{eqnarray}
    -{(1-a^{2})^{2}\over g_{c}} &=& 8 {{1+{1\over 3}a^{2}}\over{
1 - {1\over 3}a^{2}}}{\hskip 20mm} (N=2)\\
      &=& 16 (1 + a^{2}) {\hskip 20mm} (N=3)
\end{eqnarray}
they become 8.34 and 17.0 at $a=1/4$ for N=2 and N=3, respectively.

 The next order of $O(a^{4})$ is calculated for $N=2$ and approximated
by [1,2] Pad\'e,
\begin{equation}
  -{(1-a^{2})^{2}\over {g_{c}}} = 8(1 + {2\over 3}a^{2}-{1\over 15}a^{4})
 \simeq 8{{1+{31\over 15}a^{2}}\over{1+{7\over 5}a^{2}-{13\over 15}a^{4}}}
\end{equation}
which becomes 8.332 at $a=1/4$. If we estimate the exponent $\nu$ from these
small matrix result by the ratio method (10), we obtain $R_{3}=D_{3}/D_{2}
=0.795$ at a=1/4, which leads to $\nu=0.82$ according to the same estimation of
Fig.1. This value is slightly larger than the pur gravity result $\nu=0.8$.
The exact value of two matrix case is known as $\nu=6/7=0.853$.

The small $a$ expansion  without use of
the formula of (17) is also possible. It is indeed necessary to develop such
method for general n-Ising case, $2^{n}$ matrix models, since (17) is only
applied to two matrix model.
  For example, in the   N=2 case, the $2 \times 2$ matrix is represented by
\begin{equation}
      M_{2} = \left(\matrix{ c&b^{*}\cr
                             b&d\cr}\right),
\end{equation}
where we take $M_{1}=diag(\lambda_{1},\lambda_{2})$.
Then Jacobian becomes
\begin{equation}
      J = { \xi_{1} - \xi_{2} \over { \sqrt{(\xi_{1}-\xi_{2})^{2} - 4 |b|^{2}
}}}
\end{equation}
with two eigenvalues of $M_{2}$ $\xi_{1}$ and $\xi_{2}$.
 Expanding a-dependent term in the exponent, and integrating by $|b|$,
we have an effective Hamiltonian,
\begin{eqnarray}
&\int& d|b|^{2}{(\lambda_{1}-\lambda_{2})^{2}
(\xi_{1}-\xi_{2})\over{\sqrt{(\xi_{1}-\xi_{2})^{2}-4|b|^{2}}}}
\exp[{a\over 2}(\lambda_{1}+\lambda_{2})(\xi_{1}+\xi_{2}) \nonumber\\
&+&{a\over 2}(\lambda_{1}-\lambda_{2})
\times\sqrt{(\xi_{1}-\xi_{2})^{2}-4|b|^{2}}] \nonumber\\
&=& {(\lambda_{1}-\lambda_{2})^{2}(\xi_{1}-\xi_{2})^{2}\over {4}}[
1+{a^{2}\over 24}(\lambda_{1}-\lambda_{2})^{2}(\xi_{1}-\xi_{2})^{2}]
\end{eqnarray}
where we neglected the term of order $a$ which vanishes after integration of
$\lambda$, and also  we dropped
 terms which vanishes for $\lambda_{1}=-\lambda_{2}$.
We have checked that the result of (30) is obtained
 by this method up to order $a^{2}$.

  It is possible to expand the off-diagonal elements $b$
  and trancate at the sufficient order both for the measure and for
the exponent.
The diagonal elements $c$ and $d$ are given by solving the characteristic
equation in a perturbation of $|b|^{2}$,
\begin{equation}
     c = \xi_{1} - {|b|^{2}\over {\xi_{1}-\xi_{2}}} + O(|b|^{4})
\end{equation}
\begin{equation}
     d = \xi_{2} + {|b|^{2}\over {\xi_{1}-\xi_{2}}} + O(|b|^{4})
\end{equation}
 Jacobian is $J=\partial (c,d)/\partial (\xi_{1},\xi_{2})$,
\begin{equation}
   J = 1 + {2|b|^{2} \over {(\xi_{1}-\xi_{2})^{2}}} + O(|b|^{4})
\end{equation}
   After integration by these off-diagonal elements $|b|^{2}$, which is
bounded as $|b| \le (\xi_{1}-\xi_{2})/2$ , we have
an effective hamiltonian written only by the eigenvalues of two matrices.
 Taking order
$|b|^{2}$ term and up to order $a^{2}$, we obtain very close result of
(27). Including higher expansion of $|b|^{2}$, the result becomes
improved. This method can be applied for general $N\times N$ matrix.

\vskip 10mm
\begin{center}
{\large 5. n-Ising Model}
\end{center}

     We have considered Ising model on a random surface as two-matrix
model in the large N limit. The extension
of this model corresponds to the  increasing the number of species or colors
  of Ising spin. Only same color Ising spin can interact each other.
Denoting the number of colors by n, this n-Ising model is represented
by $2^{n}$ matrix model [4,5]. $2^{n}$ configulations of up and down
n-spins on
a vertex are represented by $2^{n}$ matrices.

   The Hamiltonian of this matrix model is given by multi
matrix model  similar as
two matrix model, in which $g$ is a common coupling constant of
${\rm Tr} M^{4}$ term and the coefficients of ${\rm Tr} M_{i}M_{j}$
is obtained by the inverse matrix of Boltzmann weight of spin interaction.

  In the previous works [4], we have calculated numerically the cosmological
constant $A=-1/g_{c}$, the  critical value $a_{c}$, and various critical
exponents including the string susceptibility. It is desirable, however,
to develop the method to calculate directly the cosmological constant
$-1/g_{c}$.  The main difficulty may be that there is no available formula
for the integration of matrix, and no systematic method to rewrite
the Hamiltonian only by the eigenvalues of matrices.

   We  apply the method which has been explained in the previous
section,  (38) or (42).
 Since the number of matrices increases rapidly as $2^{n}$,
we represent only n=2 and n=3 Hamiltonian here. They are described by
four matrices and eight matrices,respectively.
\begin{eqnarray}
H(n=2) &=& {1\over {2}} {\rm Tr} (M^{2}_{1}+M^{2}_{2}+M^{2}_{3}+M^{2}_{4})
  - a {\rm Tr}( M_{1}M_{2}+M_{2}M_{3}+M_{3}M_{4}+M_{1}M_{4}) \nonumber\\
  &-& a^{2}{\rm Tr}( M_{1}M_{3}+M_{2}M_{4})
   + {g\over 4}{\rm Tr} (M^{4}_{1}+M^{4}_{2}+M^{4}_{3}+M^{4}_{4})
\end{eqnarray}
\begin{eqnarray}
H(n=3) &=& {1\over 2} \sum_{i=1}^{i=8} {\rm Tr}M_{i}^{2}\nonumber\\
       &-&a{\rm Tr}(M_{1}M_{2}+M_{1}M_{3}+M_{1}M_{5}+M_{2}M_{4}+M_{2}M_{6}+
       ...)\nonumber\\
       &-&a^{2}{\rm Tr}(M_{1}M_{4}+M_{1}M_{6}+M_{1}M_{7}+M_{2}M_{3}+M_{2}M_{5}
       + ....)\nonumber\\
       &-&a^{3}{\rm Tr}(M_{1}M_{8}+M_{2}M_{7}+M_{3}M_{6}+M_{4}M_{5})\nonumber\\
       &+& {g\over N}{\rm Tr}(\sum_{i=1}^{i=8}M_{i}^{4})
\end{eqnarray}

 For N=2 case, using the representation of (37), we have for small $a$,

\begin{equation}
    -{1\over g_{c}} = {8\over{1 - {8n\over 3}a^{2}}} + O(a^{4})
\end{equation}

   For n-Ising case, the value of $-(1-a^{2})^{2n}/g_{c}$ at a=1 becomes
$2^{n}$ times of a=0 [4]. Therefore, we take [1,1] Pad\'e approximation
using this condition as
\begin{equation}
    -{(1-a^{2})^{2n}\over g_{c}} = 8 {[ 1+({{2n-3}\over 3}+{2n\over
3(2^{n}-1)})a^{2}]\over{[1+(-1+{2n\over{3(2^{n}-1)}})a^{2}]}}
\end{equation}

This Pad\'e approximation gives 8.34 for n=1 and 8.69 for n=2 at $a=1/4$.
   In the previous series analysis [4], the critical value of $-(1-a^{2})^{2n}
/g_{c}$ is estimated as 54.0 at $a_{c}=1/4$ for n=2, and 59.2 at $a_{c}=0.23$
 for n=4, and 63.8 at $a_{c}=0.21$ for n=6.

 It is important to find n dependence of $g_{c}$ for finite N matrix as
(45). We have only discussed N=2 case, and therefore we can not make
any analysis of the exponent $\nu$ at this stage.
\vskip 20mm
\begin{center}
{\large 6. Discussion}
\end{center}

   In this paper, we  have discussed the finite N saddle point equation and and
its solution. We have obtained the correct
 scaling exponent for one matrix model
and a modified one matrix model through finite N scaling.
The exponent $\nu$ has been determined by
 how finite N critical value of $g_{c}$ approaches to the $n=\infty$ critical
value $g_{c}$ (10). Our result shows that even small N, the finite size $N$
scaling works very well. This is rather remarkable and it may be
related to the fact that the small g series expansion [4] gives also correct
estimation  of $\gamma_{st}$ by small orders.
 Although we did not study the multicritical behavior
described by ${\rm Tr} M^{2l}$ terms, we believe that
  our method works
for such cases.

  For two matrix model, we
find the saddle point method is also
effective although we restricted our analysis
on small matrices. The more detailed and higher order analysis are necessary
for this case. We have shown that small $a$ expansion is useful and
Pad\'e analysis will give quite accurate numerical value for the critical
value of $g_{c}$ fora finite N.

  For n-Ising model, we have only studied very small matrices. Using small
$a$ expansion and Pad\'e approximation, we have discussed $n$ dependence of
the critical value $g_{c}$. In this case also, the approach to the $N=\infty$
critical value $g_{c}$ seems smooth, and analysis is possible. This is
consistent with the series expansion by planar diagramms [4]. There is
no tachyonic instability in this model.

   The most interesting problem of n-Ising model is to find the value of
$\gamma_{st}$ for large $n$. In the previous analysis of the perturbation
of $g$, $\gamma_{st}$ seems increasing for $c=n/2>1$.  For the large value
of $n$, we need higher order calculation in the perturbation method [4].
 Our present finite N saddle point method is
complemental one in this respect. We are planning to study more detail
for n-Ising model by making combined analysis with a series expansion..

  We have considered only $d=0$ matrix models. It is also interesting to
extend our finite N analysis for $d=1$ matrix models which becomes
quantum mechanical problems. This problem will be discussed in other place.

\begin{center}
{Acknowledgement}
\end{center}
\vskip 5mm
   The numerical calculation is supported by a Grant-in-Aid for Scientific
Research by the Ministry of Education, Science and Culture.
\newpage
\begin{center}
              {\bf References}
\end{center}
\vspace{  4mm }

\begin{description}
\item [{[1]}] V. A. Kazakov, "fields, strings and critical phenomena",
              Les Houches 1988, edited by Br\'ezin and J. Zinn-Justin,
              P.369, North-Holland, Amsterdam (1990).\\
              D. V. Boulatov and V. A. Kazakov, Phys. Lett. {\bf B187}
              (1987) 379.
\item [{[2]}] V. G. Knizhnik, A. M. Polyakov and A.A. Zamolodochikov,
              Mod. Phys. Lett. {\bf A3} (1988) 819.\\
              F. David, Mod. Phys. Lett. {\bf A3} (1988) 1651.\\
              J. Distler and H. Kawai, Nucl. Phys. {\bf B321} (1989) 509.
\item [{[3]}] E. Br\'ezin, C. Itzykson, G. Parisi and J. B. Zuber,
              Commun. Math. Phys. {\bf 59} (1978) 35.
\item [{[4]}] E. Br\'ezin and S. Hikami, Phys. Lett. {\bf B283} (1992) 203.\\
              S. Hikami and E. Br\'ezin, Phys. Lett. {\bf B295} (1992) 209.\\
              S. Hikami, Phys. Lett. {\bf B305} (1993) 327.
\item [{[5]}] C. A. Baillie and D. Johnston, Phys. Lett. {\bf B286} (1992)
44.\\
              J. Ambjorn, B. Durhuus, T. J\'onsson and G. Thorleifsson,
              Nucl. Phys. {\bf B398} (1993) 568.
\item [{[6]}] M. L. Mehta, Commun. Math. Phys. {\bf 49} (1981) 327.
\item [{[7]}] E. Br\'ezin and  V. A. Kazakov, Phys. Lett. {\bf B236} (1990)
144.\\
              D. J. Gross and A. A. Migdal, Phys. Rev. Lett. {\bf 64} (1990)
127.\\
              M. R. Douglas and S. H. Shenker, Nucl. Phys. {\bf B335} (1990)
635.
\item [{[8]}] E. Br\'ezin and J. Zinn-Justin, Phys. Lett. {\bf B288} (1992) 54.
\item [{[9]}] L. V. Lipatov, Zh. Eksp. Teor. Fiz. {\bf 72} (1977) 411.[
              Sov. Phys. JETP {\bf 45} (1977) 216.]
\item [{[10]}] M. J. Bowick and E. Br\'ezin, Phys. Lett. {\bf B268} (1991) 21.
\item [{[11]}] S. Hikami and E. Br\'ezin, J. Phys. {\bf A12} (1979) 759.
\item [{[12]}] S. Das, A. Dhar, A. Sengupta and S. Wadia, Mod. Phys. Lett.
               {\bf A5} (1990) 1041.
\item [{[13]}] C. Itzykson and J. -B. Zuber, J. Math. Phys. {\bf 21} (1980)
               411.
\end{description}
\newpage
  {\bf Table 1.}
  The critical value of $g$ obtained for one matrix model with size N.\\

\begin{tabular}{|c|c|c|c|c|c|c|c|}
\hline
N & 2& 3& 4& 5& 6& 7& 8\\
\hline
${-1/g_{c}}$ & 8& 16& 21.2& 24.8& 27.4& 29.4& 31.1\\
\hline
\end{tabular}
\vskip 30mm
  {\bf Table 2.}
   The critical value of $g$ at a fixed $g'=-9/256$ for $({\rm Tr} M^{2})^{2}$
model.\\

\begin{tabular}{|c|c|c|c|c|c|c|c|}
\hline
N & 2 & 3 & 4 & 5 & 6 & 7\\
\hline
${-1/g_{c}}$ & 11.13 & 25.60 & 35.92 & 43.29 & 48.78 & 52.99\\
\hline
\end{tabular}
\newpage
\begin{center}
{\bf Figure caption}
\end{center}
\vskip 10mm
{\bf Fig. 1.}
The ratio method for the scaling exponent $\nu$ of one matrix model.
The ratio $R_{N}=D_{N}/D_{N-1}$ is plotted against $1/(N+1)$ where
$D_{N}= 48 + 1/g_{c}$. The slope gives $\nu=4/5$.

\noindent{\bf Fig. 2.}
The critical lines obtained from  finite N saddle point solutions for
$g'({\rm Tr} M^{2})^{2}/N^{2}$ interaction. The lines are N=2,3,...,7
from the left respectively. The dotted line is the critical line for
$N=\infty$ in $g-g'$ plane. The string susceptibility $\gamma_{st}$
becomes 1/3 at $g=-3/256$ and $g'=-9/256$.

\noindent{\bf Fig. 3.}
The critical value of $-(1-a^{2})^{2}/g_{c}$ is shown for N=2,3,5,7.
The line of $N=\infty$ is exact value and it has a critical point $a=1/4$.
Two lines of N=$\infty$ are expressed by (21) and (22). The dotted lines
are the low temperature solutions.
\end{document}